\documentclass[%
superscriptaddress,
twocolumn,
amsmath,amssymb,
aps,
]{revtex4-2}

\usepackage{graphicx}
\graphicspath{{fig/}}

\usepackage[dvipsnames,table,xcdraw]{xcolor}
\definecolor{rmpblue}{HTML}{2e3092}
\usepackage[
	colorlinks=true,
	citecolor=rmpblue,
	linkcolor=rmpblue,
	filecolor=magenta,
	urlcolor=rmpblue,
]{hyperref}

\usepackage{physics}
\usepackage{upgreek}

\newcommand{\qqoute}[1]{``#1''}

\newcommand{\iu}{\mathrm{i}\mkern1mu}
\newcommand{\eu}{\mathrm{e}\mkern1mu}

\newcommand{\affilANU}{Nonlinear Physics Centre, Research School of Physics, Australian National University, ACT 2601, Australia}

\begin{document}
\title{Acoustic angular sorting of resonant subwavelength particles} 

\author{Ivan Toftul}
\email{ivan.toftul@anu.edu.au}
\affiliation{\affilANU}

\author{Yuri S. Kivshar}
\affiliation{\affilANU}

\author{Mikhail Lapine}
\email{mikhail.lapine@uts.edu.au}
\affiliation{School of Mathematical and Physical Sciences, University of Technology Sydney, NSW 2007, Australia}
\affiliation{Qingdao Innovation and Development Centre of Harbin Engineering University, Qingdao, P.R.~China}

\begin{abstract}
We suggest a dynamical mechanism for angular sorting of subwavelength particles in accord with their resonances and sizes, realised with the forces imposed by acoustic (ultrasound) waves with different wavelengths. We analyse how the acoustic force acting on a small particle depends on its size relative to the ultrasound wavelength, and how the detuning between the two different beams influences the size range and angular distribution for unambiguous  sorting outcomes for a given size range. We predict a range of scenarios depending on the particular materials and provide several feasible examples and discuss their practical realisation. 
\end{abstract}

\maketitle

\section{Introduction}

Production, functionalisation, trapping and various remote manipulation of of micro- and nanoparticles is an area of ever-growing importance, in line with the advances in nanotechnology and fabrication, microfluidics, medical and biological use of small particles, medication delivery, in-vivo and in-vitro diagnostics and therapy of many kinds
\cite{Treuel201415053,Rennick2021266,Anastasiadis2022,Wu202215627,Babicheva2024AdvOptPhotonics}.
Given that many of the high-throughput fabrication approaches typically result in a collection of particles with a significant size dispersion, various techniques of nanoparticle separation and sorting are of key importance 
\cite{Kowalczyk2011CurrOpinColloidInterfaceSci}.
Traditional separation and sorting methods like density-gradient centrifugation, chromatography or microfluidic processing
\cite{Imasaka1995AnalChem,Contado2009JChromatogrA,Mastronardi2011JACS,Xie2020ACSNano}
are however not universally applicable and are not particularly fast, but a range of  
optical \cite{Jonas2008Electrophoresis} 
and acoustic \cite{Drinkwater2016LabChip,Kolesnik2021LabChip} 
methods has been developing over the past decades.
To this end, of particular importance is the use of optical or acoustic forces 
\cite{Dholakia2020NRP,Toftul2024arXiv}
for particle manipulation, as it was first well-established in the areas of optical tweezers 
\cite{Shilkin2014JETPLett,Ren2021ACSNano,Sirotin2022BiomedOE},
and later on, owing to increasing understanding of optical forces and torques
\cite{Ng2008APL,Nieto-Vesperinas2011JOSA,Lank2018OE,Xu2020LPR,Shi2022NL},
applied towards nanoparticle sorting
\cite{Ploschner2012NL,Andres-Arroyo2016NL,ShiLyuShc17},
more typically in combination with surface fields
\cite{Marchington2008OE,Lin2012OE,Li2014ACSNano,Shilkin2022JETPLett,Shi2022PhotRes,Bulgakov2023OL}
or microfluidic techniques
\cite{MacDonald2003N,Sajeesh2014MicrofluidNanofluid,Shi2020ACSPhot,Babaliari2023}.
Similarly with the use of acoustics, microfluidic sorting approaches have been quite developed 
\cite{Wu2019MicrosystNanoeng,Hossein2023BiophysRev,Stringer2023APR,Wei2024AdvMaterTechnol}.

In the context of particle sorting, a great benefit of direct manipulation with forces coming from propagating waves, is the applicability for sorting tasks of a significant volume. While it may not be necessarily as precise as manual manipulation such as with tweezers, much faster separation processes pose a decisive advantage here. At the same time, direct sorting is much more streamlined than traditional separation techniques as it can be easily implemented directly upon production or on the application spot with almost no additional media and post-sorting purification or dialysis required. Also, direct sorting possesses further control freedom as compared to microfludics, and in particular implies a possibility to work in locked sterile or chemically isolated environments.

\begin{figure}[t]
    \centering
    \includegraphics[width=\linewidth]{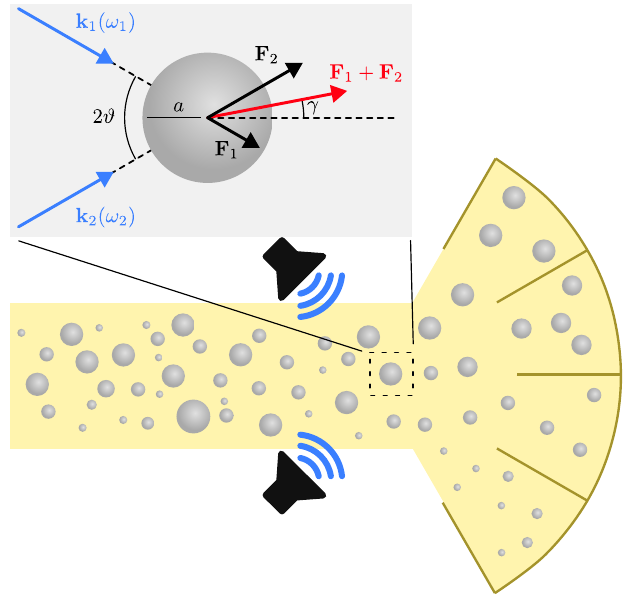}
    \caption{%
    Theoretical setup of the considered sorting scheme: 
    two waves propagating in the directions shown by wavevectors $\vb{k}_1(\omega_1)$ and $\vb{k}_2(\omega_2)$ 
    at an angle $2\theta$ to each other, 
    are incident on a particle and induce acoustic pressure forces $\vb{F}_1$ and $\vb{F}_2$, 
    having different magnitudes owing to resonant properties of the particle. 
    The resulting net force is therefore directed at some angle $\gamma$ relative to the central line,
    depending on the ratio between $F_1$ and $F_2$ for each particular particle.
    }
    \label{fig:problem}
\end{figure}

A recent extension of all-optical sorting approaches towards a simultaneous wide-range size separation was recently suggested with the use of optical pushing forces over Mie-resonant particles using two different beams at an angle to each other~\cite{ShiLyuShc17}. In this way, dielectric particles of different sizes are pushed in different directions in space and so can be collected in groups according to their size. 
At the same time, the effect of simultaneous acoustic radiation pressure with the waves of two different frequencies has been known in acoustics for almost half a century~\cite{Marston1980JASAa,Marston1980JASAb,Marston2020JQSRT,Marston2024JASA} whereby resonance effects, analogous to typical manifestations of optical Mie resonances, have been occasionally considered~\cite{Marston2007JASA,Asaki1994JASA}. However, no angular setups have been reported to date.

Now, realisation of an analogous approach for all-acoustic sorting, as shown conceptually in Fig.~\ref{fig:problem}, poses several advantages as compared to the optical implementation~\cite{Dholakia2020NRP}. First, acoustic wave sources are much cheaper and easier in exploitation compared to lasers, which promises wide-spread implementation beyond laboratory or hi-tech industry. Second, contrary to most lasers, acoustic sources are easily tunable in a wide range it terms of wavelength and intensity, which provides excellent opportunities for dynamic adjustment and reconfiguration with a single device. Third, the overall frequency range of the available ultrasound sources makes it possible to realise a sorting scheme for a huge range of particle sizes from cm-scale down to $\upmu$m-scale, although it cannot reach out down to nanoparticles where optical sorting is capable. 
Thus, remote acoustic sorting may be foreseen as a robust and industry-disposed method for particle manipulation. 
At the same time, the lack of tight focusing poses no problem for the proposed scheme, as the dimensions of the sorting area are not much greater than the wavelength, whereas coherence is not required for the success of the sorting process.

A prerequisite for the successful implementation of angular sorting is the significant understanding achieved in recent years regarding acoustic forces and torques~\cite{Silva2013JASA,Toftul2019PRL,Fan2019PRA,Sepehrirahnama2022Oct,Smagin2024PRAppl,Toftul2024arXiv}. 
At the same time, acoustic Mie resonances have received considerable attention in the literature~\cite{Lu2017APL,Zhang2018SR} including superscattering effects~\cite{Liu2019PRA}, however insofar they have not been assessed in the context of acoustic manipulation.

We will therefore provide a detailed parametric analysis permitting angular acoustic sorting of particles according to their size and material properties, including the analyses of the Mie resonance effects on the acoustic pressure force.

In order to make analytical and numerical calculations straightforward, we will assume sorted particles to be spherical, 
so that their resonances as well as resulting forces can be analytically calculated. 
For a practical collection of realistic particles, our  approach still works, whereby actual sorting will be achieved according to resonances and can be still calibrated suitably for a specific purpose; but here we will only concentrate on general and conceptual findings.

\section{Acoustic pressure for sorting}

For the purposes of this paper, we can consider particles with sizes sufficiently smaller than the width of acoustic beams used for sorting, so that the beams can be treated as plane waves.

We will deal with monochromatic acoustic fields of frequency $\omega$ in a homogeneous dense medium (fluid or gas) assuming $\eu^{-\iu \omega t}$ time dependence. The complex pressure and velocity fields, $p(\mathbf{r})$ and $\mathbf{v}(\mathbf{r})$, obey the master wave equations~\cite{landau2013fluid}
\begin{equation}
\iu \omega \beta p = \nabla \cdot \mathbf{v}, \qquad   \iu \omega \rho \mathbf{v} = \nabla p,
\end{equation}
where the medium is characterised by two parameters: compressibility $\beta$ and mass density $\rho$; the velocity of sound is then $c_s = 1/\sqrt{\rho \beta}$.

For a spherical particle of radius $a$, made of material with mass density $\rho_p$ and compressibility $\beta_p$,
immersed in a host medium described by compressibility $\beta$ and mass density $\rho$,
an incident acoustic plane wave $p = p_0\eu^{\iu \vb{k} \cdot \vb{r}}$ imposes the following acoustic pressure force
~\cite{Hasegawa1977JAcoustSocAm,Yosioka1955}:
\begin{equation}
    \vb{F}^{\text{pres}} = \bar{\vb{k}} \, {p_0^2 \beta  \sigma^{\text{pres}}}/{2} ,
    \label{eq:Fpr}
\end{equation}
where $\bar{\vb{k}} = \vb{k}/k$, $k = |\vb{k}| = \omega \sqrt{\beta \rho}$ is the wavenumber in the host media, 
and \textit{pressure cross section} $\sigma^{\text{pres}}$ is given by
\begin{align}\label{eq:sigmapr}
    &\sigma^{\text{pres}} = \sum \limits_{n=0}^{\infty} \sigma^{\text{pres}}_n\\ 
    &=- \frac{4\pi}{k^2} \sum \limits_{n=0}^{\infty} \bigg[ (2n+1) \Re \left(a_n\right) + 2(n+1) \Re\left( a_n^* a_{n+1} \right) \bigg]. \nonumber
\end{align}
where $a_n$ are the acoustic Mie coefficients~\cite{Toftul2019PRL,blackstock2000fundamentals,Yosioka1955}
\begin{equation}
    a_{n} = \frac{ \sqrt{\bar{\beta}}\, j_n^{\prime} (k_p a) j_n(ka) - \sqrt{\bar{\rho}}\, j_n(k_p a) j_n^{\prime}(ka)}{ \sqrt{\bar{\rho}}\, j_n(k_{p} a) h_n^{(1)\prime} (ka) - \sqrt{\bar{\beta}}\, j_n^{\prime}(k_{p} a) h_n^{(1)}(ka)},
    \label{eq:an}
\end{equation}
where $j_n$ is the spherical Bessel function, $h_n^{(1)}$ is the spherical Hankel function of the first kind, the prime denotes derivative with respect to the argument, $\bar{\rho} = \rho_p /\rho$ and $\bar{\beta} = \beta_p /\beta$ are the relative density and compressibility, and  $k_p = \omega \sqrt{\beta_p \rho_p} = k \sqrt{\bar{\beta} \bar{\rho}}$ is the wavenumber inside the particle. Thus, scattering coefficient \eqref{eq:an} is the function of only three parameters: $a_n(ka, \bar{\rho},\bar{\beta})$.

Let us further analyse Eqs.~\eqref{eq:Fpr}--\eqref{eq:sigmapr}. 
Terms $\propto \Re(a_n)$ are the acoustic radiation pressure forces associated with the \textit{phase gradients} of the incident field, while terms $\propto \Re\left( a_n^* a_{n+1} \right)$ are \textit{recoil} forces associated with the momentum carried away by the \textit{scattered} wave field~\cite{Toftul2024arXiv}.
By introducing \textit{Mie angle} $\phi_n$ as $a_n = |a_n(\phi_n)| \eu^{\iu \phi_n}$~\cite{Rahimzadegan2020OE,Hulst1981},  we find the general form of the Mie scattering coefficient based on energy conservation  $\Re(a_n) + |a_n|^2 = 0$ (which is equivalent to the vanishing partial absorption cross  section)~\cite{Toftul2024arXiv,Zhang2011JASA} to be $a_n = - \cos \phi_n \eu^{\iu \phi_n}$, $\phi_n \in (-\pi/2, \pi/2)$. 
The Mie angle formulation is equivalent to the well-known representation of scattering coefficients in a form $a_n = (1/2)(e^{2\iu\eta_n} - 1)$, where $\eta_n$ is the complex valued wave phase shift~\cite{Zhang2016JASA,Marston2017JASA,Marston2024JASAEL}. This approach has proven to be a particularly useful in low-frequency expansions and in analyzing radiation forces. In that formalism, the condition of vanishing absorption corresponds to real $\eta_n$, and energy dissipation can be introduced by the complex part of $\eta_n$~\cite{Zhang2016JASA}. 
Either approach implies the following limits:
\begin{equation}\label{eq:limits}
  \begin{aligned}
    -1 \leqslant \Re( a_n) \leqslant 0 \, ,   \\
    -\frac{1}{8} \leqslant \Re( a^{*}_n a_{n+1}) \leqslant 1 \, .
  \end{aligned}  
\end{equation}
Maximisation of the recoil terms occurs at the so-called generalised Kerker condition, which corresponds to the maximised forward scattering~\cite{Toftul2024arXiv,Wu2021APE,Wei2020NJP}. 
However, the partial pressure cross section $\sigma^{\text{pres}}_n$ in Eq.~\eqref{eq:sigmapr} gets maximised when, at the same time, $\Re( a_n) \to -1$ as well as $\Re( a^{*}_n a_{n+1}) \to - 1/8$ (an anti-Kerker condition).
Equations \eqref{eq:limits} define the partial \textit{acoustic pressure force limits}. Similar limits were found for the optical forces in Ref.~\cite{Rahimzadegan2017PRB}.

A couple of remarks regarding dissipation must be made:
\begin{itemize}
    \item[(i)] In the current formalism, the imaginary parts of the relative material parameters,  $\Im \bar{\beta}$ and $\Im \bar{\rho}$, correspond to the dissipative losses, such as those caused by porosity of the material~\cite{Umnova2005AA};
    \item[(ii)] Additionally, we assume that the viscous Stokes layer thickness is much smaller than the particle radius, which justifies neglecting the frequency-dependent boundary layer contributions to dissipation which otherwise become significant~\cite{Zhang2014JASA}.
\end{itemize}

\section{General analysis of angular sorting scheme}

The core idea behind angular sorting in acoustic is the use of a pair of sound pressure forces arising over resonant microparticles at different sound frequencies, $\omega_1$ and $\omega_2$. A conceptual schematic of this approach is illustrated earlier in Fig.~\ref{fig:problem}.
Two plane waves of different frequencies, propagating in different directions with wavevectors $\mathbf{k}_1(\omega_1)$ and $\mathbf{k}_2 (\omega_2)$ with opening angle $2\vartheta$, impose pressure forces $\vb{F}_1$ and $\vb{F}_2$ over a microparticle. 
The magnitude of these forces is determined by a resonance response of the microparticle, essentially Mie-type resonances which we would like it to have in the frequency range of interest. 
The resulting total force is thereby directed at some angle $\gamma$ relative to the ``default'' median direction (corresponding to the absence of forces or equal forces).
The resonances of particles are determined by its size, shape, constituent material as well as surrounding medium.
The presented setup therefore prompts for a rich parametric analysis and a wide range of different sorting scenarios.

The primary step in our analysis is the general evaluation of how the balance between resonant forces depends on the wavelength of incident sound beams and on the difference between their frequencies. 

Given two different waves ($k_{1,2}$), imposing two different acoustic pressure forces \eqref{eq:Fpr} with magnitudes $F_{1,2}$, the eventual deflection angle $\gamma$ relative to the non-perturbed median line (Fig.~\ref{fig:problem}) can be calculated as
\begin{equation}
    \gamma = \arctan \left( \frac{F_2 - F_1}{ F_2 + F_2} \tan \vartheta \right).
\end{equation}
We assume a negligible contribution from the interference of bichromatic (or polyphonic) terms oscillating at the frequency $\Delta \omega = c_s (k_1 - k_2)$, thus any related effects fall outside of this work~\cite{Silva2006PRL,Morrell2024PRE}

\begin{figure}[t]
    \centering
    \includegraphics[width=\linewidth]{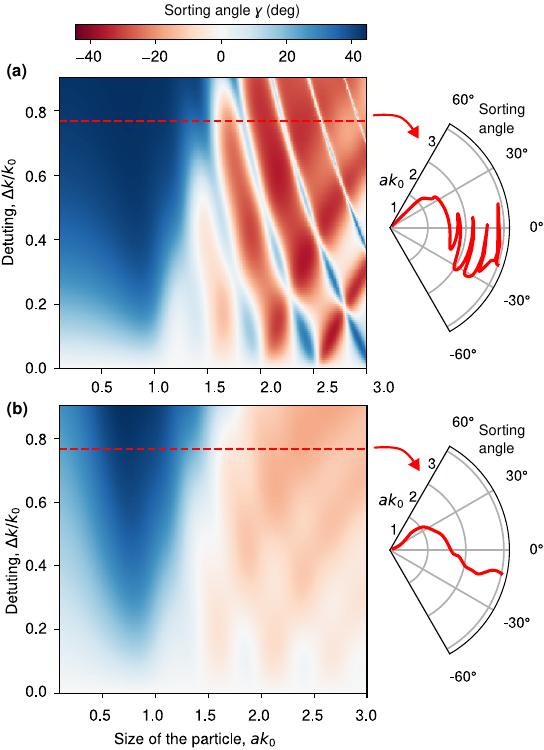}
    \caption{Concept of the dynamical angular sorting. The parametric map for the angle for the declination angle for lossless (a) and lossy (b) resonant acoustic particles. Relatives parameters of the sphere is $\bar{\rho} = 3 (+0.3\iu)$ and $\bar{\beta} = 2(+0.2\iu)$.}
    \label{fig:loss_vs_lossless}
\end{figure}

For the ease of parametric analysis, we introduce two dimensionless quantities: (i) the relative difference between the frequency of the two waves, characterised by relative detuning parameter $\Delta k/k_{0} = (k_2 - k_1)/k_{0}$ and (ii) size parameter of the sphere $a k_{0}$, where $k_{0} = (k_1 + k_2)/2$ is the average wavenumber. The sorting angle thus becomes a function of theses two parameters: $\gamma = \gamma \left( \Delta k/k_0, a k_{0} \right)$.

\begin{table*}
\caption{Acoustic properties of various materials.
}
\label{t:materials}
\begin{ruledtabular}
\begin{tabular}{lllll}
Material &
Mass density, $\rho$~[kg/m$^3$] &
Speed of sound, $c_s$~[m/s]&
Compressibility, $\beta$~[m$^2$/N] & 
Losses, $\delta$~[s/m]\\
\colrule
Air~\cite{Wong1986JAcoustSocAm} & 1.28 & 343 & $6.6\times 10^{-6}$ & $\sim 3\times 10^{-7}$ @ 30~kHz \\
Water~\cite{DelGrosso1972JAcoustSocAm} & 997 & 1500 & $4.5\times 10^{-10}$ & $\sim 3\times 10^{-9}$ @ 30~kHz \\
pSi~\cite{Ba2017SciRep}  & 1040 & 39 & $6.3\times 10^{-7}$ & $\sim 1 \times 10^{-3}$ @ 30~kHz \\
Aerogel~\cite{Mazrouei-Sebdani2021JNonCrystSolids,Moretti2017EnergyProcedia} & 70 & 100 & $1.4\times 10^{-6}$ & $\sim 1 \times 10^{-3}$ @ 1.5~kHz\\
SF$_6$~\cite{Zuckerwar2002} & 6.17 & 133 & $9.2\times 10^{-6}$ & $\sim 1 \times 10^{-4}$ @ 1.5~kHz
\end{tabular}
\end{ruledtabular}
\end{table*}

In Fig.~\ref{fig:loss_vs_lossless} we depict the deflection angle $\gamma$ (shown with the color map), depending on the relative size of the particle $a k_{0}$ (horizontal axis), and on the relative detuning $\Delta k / k_{0}$ (vertical axis).
For this generic illustrative example we consider a three-fold contrast in density between the particle material and the host medium, $\bar{\rho} = 3$, and a two-fold contrast in compressibility $\bar{\beta} = 2$.
The upper figure presents an ideal situation with no acoustic losses, which highlights the resonances sharply, 
whereas the lower picture considers the same situation with noticeable dissipation: 
$\bar{\rho} = 3+0.3\iu$ and $\bar{\beta} = 2+0.2\iu$. 
The right-hand sides of Fig.~\ref{fig:loss_vs_lossless}\,(a),(b) show a particular example of angular distribution of resonant particles according to their size at a given frequency offset $\Delta k /k_0$, marked by red dashed lines in the left-hand plots. In other words, these polar  plots show the resulting drift direction (shown as the actual angle) for a range of particles depending on their size (indicated on the radial axis).

For a lossless case, resonances are expected to be narrow, implying very high frequency sensitivity, so for a given frequency drift direction is very sensitive to particle size. Thereby low loss increases the range of deflection angles but also contributes to sorting ambiguity whereby a few different sizes are sent in the same direction which may require further processing.
On the other hand, the presence of losses flattens the resonances, which reduces the angular span somewhat, however helps to resolve much of the ambiguity, improving reliability of angular separation.

With the above general conclusions on the operational principles of angular sorting scheme at hand, we can now proceed towards a series of examples driven by real material parameters.

\section{Sorting efficency depending on material parameters}

The strength of acoustic Mie resonances and their suitability for sorting is determined by material combinations.
Eventual sorting scenario depends on the sharpness (and the very existence) of resonances and their distribution across frequencies. 
Below, we provide several illustrative examples based on practically available material characteristics.

In Table~\ref{t:materials} densities, compressibilities and sound velocities for several materials are listed, 
whereby the velocity of sound is related to the other two characteristics as $c_s = 1/\sqrt{\beta \rho}$. 
Acoustic wave propagating in a homogeneous medium is attenuated due to the dissipation. Intensity of the sound wave decays exponentially, which is commonly described by Bugger's law as $I = I_0 \eu^{-\alpha x}$, where $x$ is the penetration depth of the acoustic wave, $\alpha$ is attenuation coefficient, and $I_0$ is the initial value of the sound intensity. 
Values of $\alpha$ are usually known from the literature for many materials. 
However, in the context of sound scattering by particles comparable with the wavelength imaginary part of the acoustic refractive index is much more useful. 
Therefore, in Table~\ref{t:materials} we also present an approximate value of losses, $\delta = \Im \left(\sqrt{\rho \beta}\right)$~[s/m], in the last column of the table, which is connected to the attenuation coefficient $\alpha$ as $\delta = \alpha/(2 \omega)$.
The relative losses $\bar{\delta}$ of a particle suspended into the host media with $c_{s,\text{host}}$  is simply $\bar{\delta} \simeq \delta c_{s,\text{host}}$ (imaginary part of the acoustic refractive index). 
For most materials, the exact origin of dissipation, which can be theoretically described as the imaginary parts of the compressibility and/or imaginary part of mass density, is actually not known~\cite{Umnova2005AA}.
As a fair approximation for $\Im(\rho)$ and $\Re(\beta)$, we assume that $\beta \to \beta (1 + \iu \epsilon)$ and $\rho \to \rho (1 + \iu \epsilon)$, so that by comparing $\delta = \Im\left(\sqrt{\rho \beta}\right)$ with the experimental data, we retrieve the required correction to the material parameters as $\epsilon \simeq \delta / \Re\sqrt{\rho \beta }$.

Efficiency of resonant sorting is determined by the pressure %
cross-section \eqref{eq:sigmapr} of a particle at resonance, as well as by the pattern of resonances: their multiplicity and sharpness. 
The resonant response is defined by the eigenmodes of the particle. In the case of a sphere this is defined by zeros of the denominator of the Mie scattering coefficients \eqref{eq:an}. %
The corresponding dependence can be explicitly represented by equation:
\begin{equation}
    \sqrt{\bar{\rho}}\, j_n(k_{p} a) h_n^{(1)\prime} (ka) = \sqrt{\bar{\beta}}\, j_n^{\prime}(k_{p} a) h_n^{(1)}(ka),
    \label{eq:an_den}
\end{equation}
where $k_p = \sqrt{\bar{\rho} \bar{\beta}} k$. 

Solutions to this equation are complex dimensionless frequencies $(ka)_{\text{res}} = \bar{\omega}_{\text{res}} - \iu \bar{\gamma}$. 
The corresponding $Q$-factor of each resonance is equal to $Q = \bar{\omega}_{\text{res}} / 2 \bar{\gamma}$. 
For the monopole term ($n=0$) the dispersion equation \eqref{eq:an_den} simplifies to
\begin{equation}
    \frac{k_p a \,  \sin(k_p a)}{k_p a \, \cos(k_p a) - \sin(k_p a)} = \frac{\sqrt{\bar{\beta} / \bar{\rho}} \,  ka}{\iu ka - 1}.
\end{equation}

\begin{figure}
    \centering
    \includegraphics[width=0.97\linewidth]{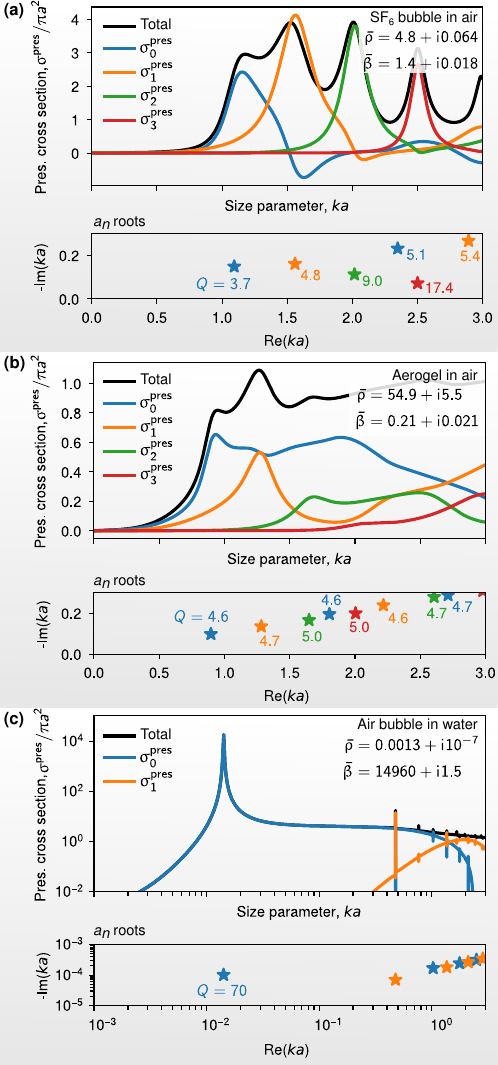}
    \caption{Three distinct examples of the normalized pressure cross sections: (a) $\mathrm{SF}_6$ bubbles in air (surface tension of the bubble is neglected), (b) aerogel spheres in air, and (c) air bubble in water. Lower panels show roots of the dispersion equation \eqref{eq:an_den}, and corresponding $Q$-factors.}
    \label{fig:pres_cross_section_and_roots}
\end{figure}

The resonance conditions can be quite nontrivial for acoustic resonances.
In Fig.~\ref{fig:pres_cross_section_and_roots}, we plot pressure cross sections normalized by the geometrical cross section $\pi a^2$.
The bottom of each panel shows the roots of Eq.~\eqref{eq:an_den} for different $n$, with the corresponding $Q$-factors indicated next to each root.
We showcase several illustrative examples: an SF$_6$ bubble in air, neglecting the bubble surface tension [Fig.~\ref{fig:pres_cross_section_and_roots}(a)]; a spherical aerogel particle in air [Fig.~\ref{fig:pres_cross_section_and_roots}(b)]; and an air bubble in water [Fig.~\ref{fig:pres_cross_section_and_roots}(c)].
One can see that the acoustic refractive index, $\sqrt{\bar{\rho} \bar{\beta}}$, is quite similar for all three cases: approximately $2.6$ for (a), $3.4$ for (b), and $4.4$ for (c). Yet, the resonant behavior is drastically different.
The most distinct case is gas bubbles in liquids, which are typically characterized by extremely high relative compressibility, $\bar{\beta} \gg 1$, and low relative mass density, $\bar{\rho} \ll 1$.
Within this range, particles exhibit a pronounced peak originating from the first monopole resonance (or the first ``breathing'' mode, which corresponds to periodic homogeneous compression and expansion of the particle~\cite{blackstock2000fundamentals,williams1999FourierAcousticsSound}).
Such a sharp monopole peak could potentially enable highly selective, narrow-band bubble sorting, as illustrated in Fig.~\ref{fig:problem}.
However, for this example we must emphasise that it is mostly provided for the sake of parametric completeness, but in practice $Q$-factor of this monopole resonance may be significantly reduced due to viscous and thermal energy dissipation, which are highly relevant for gas bubbles in liquids~\cite{Asaki1994JASA}. Apart from that, gas bubbles are also highly deformable and may undergo significant nonlinear oscillations with the applied acoustic intensities. However, for the current illustrative parametric analysis we do not account for these effects.

Low values of $\bar{\rho}$ suppress all other resonances apart from monopole. 
Furthermore, Taylor series of the monopole Mie coefficient \eqref{eq:an} for $ka \ll 1$ gives $a_0 \simeq \frac{\iu}{3} \left(\bar{\beta} - 1 \right) (ka)^3$, which means that monopole resonance does not occur in Rayleigh approximation. 
Below, we show that the symmetric case of $\bar{\rho} \gg 1$ and $\bar{\beta} \ll 1$ does not produce a single pronounced peak in the range of \textit{natural} materials, $\bar{\rho}>0$ and $\bar{\beta} > 0$. The acoustic meta-materials are still to be further studied in the aspect of acoustic forces and torques, yet many works have already showed its high potential~\cite{Li2004PRE,Nicolaou2012NatMater,Farhat2016APA,Bliokh2019PRB,Ambati2007PRB,Park2011PRL,Cselyuszka2014PhysLettA,Fok2011PRB,Lee2009PhysLettA,Huang2009IntJEngSci,Lee2016PRB,Fan2016PRB}.

Next, spheres made of aerogel have a good combination of $\bar{\beta}$ and $\bar{\rho}$ to provide resonant behavior. Unfortunately, all the advantages are outweighed by its high dissipative losses, as aerogel is a highly porous material~\cite{Mazrouei-Sebdani2021JNonCrystSolids,Moretti2017EnergyProcedia}. Thus the resulting pressure cross section does not exhibit strong resonant behavior, Fig.~\ref{fig:pres_cross_section_and_roots}(b).

A smoother spectral picture is shown for $\mathrm{SF}_6$ bubbles in air, Fig.~\ref{fig:pres_cross_section_and_roots}(a); however, we acknowledge that this is rather an artificial example from a practical perspective. Nonetheless, it serves as a good model for theoretical analysis and for illustrating resonant features.

Based on the above examples, we may conclude that high density contrast for soft particles promotes multi-resonance response and enhances sorting efficiency.

Figure~\ref{fig:pres_cross_section_and_roots} also shows \textit{partial} contributions $\sigma^{\text{pres}}_n$ to the pressure cross section \eqref{eq:sigmapr} and hence to the acoustic pressure force \eqref{eq:Fpr}. One can see that these partial contributions can be \textit{negative} in some spectral areas, while the total $\sigma^{\text{pres}}$ remains positive for passive particles in a plane wave. Negative values occur when the Kerker condition $\Re (a^*_n a_{n+1}) \approx 1$ is met together with low values of $\Re (a_n) \approx 0$ [see the limits in Eq.~\eqref{eq:limits}]. 
Configurations with complex beam shapes can produce negative or \qqoute{\textit{pulling}} acoustic forces identified in~\cite{Marston2006JASA}, which has been then explored in more detail for complex beam shapes in various acoustic~\cite{Zhang2011PRE,Xu2012EPL,Demore2014PRL,Mitri2015EPL,Mitri2015JAP,Meng2020PRA,Wang2021PRA} and optical (see the references in~\cite{Toftul2024arXiv}) scenarios. However, extending the current analysis towards such interesting effects is beyond the scope of the present paper.

\begin{figure*}
    \centering
    \includegraphics[width=0.95\linewidth]{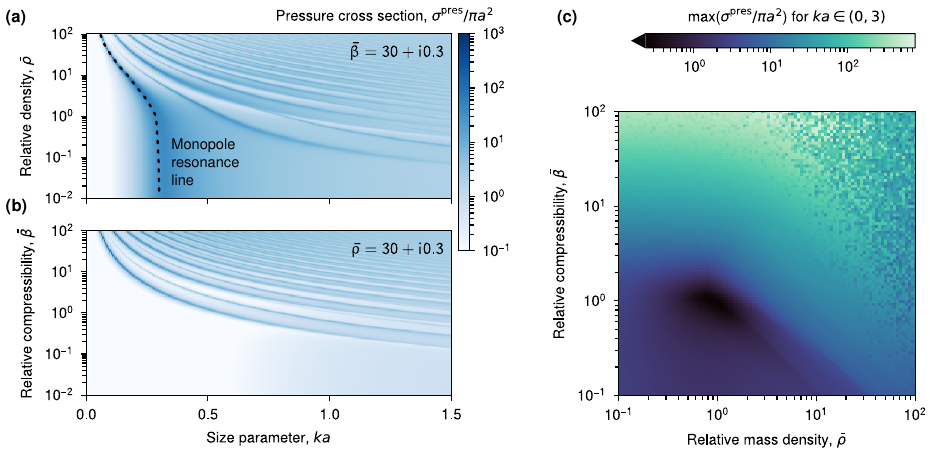}
    \caption{Normalized pressure cross section for fixed $\bar{\beta}$ (a) or for fixed $\bar{\rho}$ (b). Panel (c) shows maximum peak value in the range $ka \in (0, 3)$ of the normalized pressure cross section, $\sigma^{\text{pres}}/\pi a^2$ for different values of relative compressibility and density. Losses added by the rule $\bar{\rho} \to \bar{\rho} (1 + \iu \epsilon)$ and  $\bar{\beta} \to \bar{\beta} (1 + \iu \epsilon)$ with $\epsilon = 0.01$.}
    \label{fig:sigma_pr_map}
\end{figure*}

Another type of parametric research we would like to present, is the analysis all possible natural material combinations ($\bar{\rho}>0$ and $\bar{\beta} > 0$) in terms of two further parametric spaces.
First, we fix relative compressibility to $\bar{\beta} = 30 + \iu 0.3$, while vary size parameter $ka$ and relative density values $\bar{\rho}$, Fig.~\ref{fig:sigma_pr_map}(a). The losses are added by the rule $\bar{\rho} \to \bar{\rho} (1 + \iu 0.01)$.
One can see that for simultaneous high values of material parameters, particle must exhibit sharp resonant behavior. Surprisingly, the monopole resonance line is not suppressed even by extremely low relative mass density (the limit of \textit{pressure release} spheres, see~\cite[Sec.~6.10.2]{williams1999FourierAcousticsSound}).
Second, we fix relative mass density to $\bar{\rho} = 30 + \iu 0.3$ and vary relative compressibility values, Fig.~\ref{fig:sigma_pr_map}(b). From this parametric map we see that $\bar{\beta} \ll 1$ suppress any resonant features, regardless high values of relative mass density (the limit of \textit{rigid} spheres, see~\cite[Sec.~6.10.3]{williams1999FourierAcousticsSound}). 

It is important to point out that 
acoustic resonant scattering is \textit{asymmetric} with respect to $\bar{\rho} \leftrightarrows \bar{\beta}$ exchange; the monopole resonance plays a special role in this outcome. 
This is in contrast to optics, where situation is symmetric with the exchange between permittivity %
and permeability. %

The other key feature, general \qqoute{sharpness} of the resonances, can be also presented on a parametric plot, and for this purpose we calculate the maximum achievable pressure %
cross-section (normalised to geometric cross-section of a particle) for the size span of $0 < ka < 3$,
across the entire spectral range where any resonances are observed (if at all).
The result is plotted using a colour map in Fig.~\ref{fig:sigma_pr_map}(c) 
depending on relative compressibility and on relative density.
Apart from the obvious minimum for the zero contrast case $\bar{\beta} = \bar{\rho} = 1$, the following observations can be made from this analysis.
High compressibility contrast is most crucial for promising a possibility of efficient sorting: as we can see in the figure, pressure cross-section is strongly increased when the inclusion is much softer then the host material (such as air bubbles in water). Vice versa, for hard inclusions pressure cross-section is rather low;
Density contrast is also helpful but to a smaller extent; still it may lead to suitable outcomes even if the compressibility contrast is low (such as aerogel particles in air). 
This said, we admit that here we only provide some very general indications on the likely possibilities, as the actual pattern of resonances cannot be deduced from this parametric plot alone.

\section{Conclusions}

We have provided a theoretical parametric analysis, evaluating possibilities for angular acoustic sorting of particles according to their size, or, strictly speaking, according to frequency dependence of resonances in their pressure cross-section.
Angular sorting can be achieved by using two acoustic waves with different frequencies and we have shown that, 
by independently selecting the frequency difference between the two waves, and their average frequency, a wide parametric range is achieved, permitting to accommodate a great variety of sorting scenarios. 
We have also analysed the effect of dissipation and show that it does not limit sorting possibilities, but in fact helps to resolve potential ambiguities which arise with very sharp resonances.

Further on, we presented a general parametric analysis trying to assess an overall figure of merit for angular sorting scheme, depending on compressibility and density contrasts between the sorted particles and host medium.
The maximum achievable pressure cross-section, which we have used for that purpose, does not alone provide a complete systematic picture, but helps reaching some general conclusions such as that a high contrast in compressibility is the most desirable feature in the sorting setup, whereby soft particles in a rigid medium are most readily sorted.
This observation makes sorting schemes in water or other liquid medium highly promising.

Overall, the span of the presented parametric picture should help to select the most efficient scheme for a specific sorting scenario arising in practice. 
To this end, we should note that the actual pattern of acoustic resonances depends on the individual material combination. 
For any practical scenario, it can be analysed using the methodology described in this paper: even though exact position of resonances can be different in real particles, our separation approach does not rely on targeting specific resonances, but it exploits the inherent diversity of resonant responses across a population of particles.

Finally, we would like to reiterate on the limitations of the analysis presented in this paper:
\begin{enumerate}
    \item[(i)] Our scenario is not optimal for separation of tightly packed aggregates, so initial segregation of clusters into more diluted collection would be required.
    \item[(ii)] Additional sources of loss might affect calculations results, e.g. accurate consideration of viscous Stokes layer is essential for small particles.
\end{enumerate}
While our approach clearly has its own limitations, we are open for collaboration regarding any specific situations of interest which may arise in scientific and industrial community.

\begin{acknowledgements}
The authors thank Mihail Petrov for fruitful discussions. This work was partially supported by the Strategic Fund of the Australian National University and the Australian Research Council (grant no. DP210101292).
\end{acknowledgements}

\appendix

\section{Acoustic Mie scattering problem}

\begin{figure}
    \centering
    \includegraphics[width=0.9\linewidth]{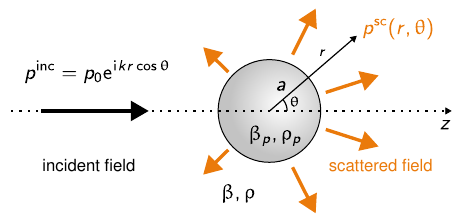}
    \caption{Mie scattering problem in acoustics. Incident plane wave is being scattered by a isotorpic sphere described by compressibility $\beta_p$  and mass density $\rho_p$  of radius $a$.}
    \label{fig:mie_scattering}
\end{figure}

We consider an isotropic particle with the compressibility $\beta_p$  and mass density $\rho_p$, and the radius $a$, located at $\vb{r} = \vb{0}$ in a homogeneous medium with parameters $\beta$ and $\rho$ (Fig.~\ref{fig:mie_scattering}). Using spherical coordinates $(r, \theta, \phi)$, the incident $z$-propagating
plane-wave field can be written as~\cite{Zhang2016JASA,Toftul2019PRL,blackstock2000fundamentals,williams1999FourierAcousticsSound}
\begin{equation}
    p^{\text{inc}} = p_0 \eu^{\iu kr \cos \theta} = \sum \limits_{n=0}^{\infty} p_n j_n(kr) P_n(\cos \theta),
    \label{eq:pinc}
\end{equation}
where $p_n = p_0 \iu^{n} (2n + 1)$, $j_n$ are the spherical Bessel functions of the first kind, and $P_n$ are the Legendre polynomials. 
Taking into account the azimuthal symmetry of the problem, the field inside the spherical particles and scattered field
outside the particles can be written as:
\begin{align}
\begin{split}
    p^{\text{par}} &= \sum \limits_{n=0}^{\infty} p_n c_n j_n(k_p r) P_n(\cos \theta), \\
    p^{\text{sc}} &= \sum \limits_{n=0}^{\infty} p_n a_n h_{n}^{(1)}(kr) P_n(\cos \theta),
\end{split} \label{eq:psc}
\end{align}
where $k_p = \sqrt{\bar{\rho} \bar{\beta}} k$ is the wave number inside the particle, and $h_n^{(1)}$ are the spherical Hankel functions of the first kind. 

The coefficients $a_n$ and $c_n$ in Eqs.~\eqref{eq:psc} should be determined from the boundary conditions, i.e., the continuity of the pressure and normal velocity component at the interface $r = a$. Using $\vb{v} = (\iu \omega \rho)^{-1} \grad p$, we have:
\begin{equation}
        p^{\text{inc}} + p^{\text{sc}} = p^{\text{par}}, \quad 
        \frac{1}{\rho} \left( \partial_r p^{\text{inc}} + \partial_r p^{\text{sc}} \right) = \frac{1}{\rho_p} \partial_r p^{\text{par}}.
        \label{eq:bc}
\end{equation}
Substituting fields \eqref{eq:pinc} and \eqref{eq:psc} into the boundary conditions \eqref{eq:bc}, we derive:
\begin{align}
    \begin{split}
        c_{n} &= \frac{ \iu \sqrt{\bar{\rho}}  / (ka)^2}{ \sqrt{\bar{\rho}}\, j_n(k_{p} a) h_n^{(1)\prime} (ka) - \sqrt{\bar{\beta}}\, j_n^{\prime}(k_{p} a) h_n^{(1)}(ka)}, \\
        a_{n} &= \frac{ \sqrt{\bar{\beta}}\, j_n^{\prime} (k_p a) j_n(ka) - \sqrt{\bar{\rho}}\, j_n(k_p a) j_n^{\prime}(ka)}{ \sqrt{\bar{\rho}}\, j_n(k_{p} a) h_n^{(1)\prime} (ka) - \sqrt{\bar{\beta}}\, j_n^{\prime}(k_{p} a) h_n^{(1)}(ka)},
    \end{split}
\end{align}
where $j_n$ is the spherical Bessel function, $h_n^{(1)}$ is the spherical Hankel function of the first kind, the prime denotes derivative with respect to the argument, $\bar{\rho} = \rho_p /\rho$ and $\bar{\beta} = \beta_p /\beta$ are the relative density and compressibility, and  $k_p = \omega \sqrt{\beta_p \rho_p} = k \sqrt{\bar{\beta} \bar{\rho}}$ is the wavenumber inside the particle.

\section{Acoustic pressure force}

The acoustic radiation force acting on a scattering particle can be calculated using the momentum 
momentum flux through a closed surface $\Sigma$ enclosing the particle~\cite{gorkov1962, Westervelt1951JASA, Toftul2019PRL,Toftul2024arXiv}:
\begin{equation}
    \vb{F} = - \oint \limits_{\Sigma} \left[  \frac{\rho}{2} \Re\left(  \vb{v}^* \otimes \vb{v}\right) + \hat{\vb{I}}\frac{1}{4} \left( \beta |p|^2 - \rho |\vb{v}|^2 \right) \right] \cdot \dd \vb{\Sigma},
    \label{eq:force}
\end{equation}
where ``$\otimes$'' shows the tensor product of two vectors, and $\hat{\vb{I}}$ is the unitary matrix. 
Here, the acoustic wave field is the sum of the incident and scattered fields outside the particle: $p = p^{\text{inc}} + p^{\text{sc}}$, $
\vb{v} = \vb{v}^{\text{inc}} + \vb{v}^{\text{sc}}$. Direct substitution of the \eqref{eq:pinc} and \eqref{eq:psc} into force \eqref{eq:force} provides an expression for the acoustic radiation pressure force from a plane wave expressed via scattering coefficients~\cite{Hasegawa1977JAcoustSocAm,Yosioka1955}:
\begin{equation}
    {F}_z^{\text{pres}} = {p_0^2 \beta  \sigma^{\text{pres}}}/{2} ,
\end{equation}
where \textit{pressure cross section} $\sigma^{\text{pres}}$ is given by
\begin{equation}
\resizebox{\linewidth}{!}{$
    \sigma^{\text{pres}} = - \dfrac{4\pi}{k^2} \sum \limits_{n=0}^{\infty} \bigg[ (2n+1) \Re \left(a_n\right) + 2(n+1) \Re\left( a_n^* a_{n+1} \right) \bigg]. 
$}
\end{equation}

\bibliography{acoustic_sorting_refs}

\end{document}